\documentclass[fleqn,usenatbib]{mnras}
\usepackage{newtxtext,newtxmath}
\usepackage[T1]{fontenc}
\usepackage{ae,aecompl}
\usepackage{graphicx}	
\usepackage{amsmath}	
\usepackage{amssymb}	
\usepackage{marvosym}
\usepackage{hyperref}
\usepackage{soul}
\usepackage{multicol}
\usepackage{booktabs}
\usepackage[table,xcdraw]{xcolor}
\usepackage{eurosym}
\newcommand{\beq}[1]{\begin{equation}\label{#1}}
\newcommand{\eeq}{\end{equation}}
\newcommand{\sub}[1]{_{\rm #1}}
\newcommand{\beqn}{\begin{equation}}
\newcommand{\eeqn}{\end{equation}}

\newcommand{\am}{a\sub{m}}
\newcommand{\ap}{a\sub{p}}

\newcommand{\Rh}{R\sub{Hill}}
\newcommand{\Rp}{R\sub{p}}

\newcommand{\Der}{\mathrm{d}}

\newcommand{\aploon}{a\sub{pl}}

\newcommand{\Teq}{T\sub{eq}}
\newcommand{\SigmaSB}{\sigma\sub{SB}}

\newcommand{\rploon}{r\sub{pl}}
\newcommand{\mploon}{m\sub{pl}}
\newcommand{\Lstar}{L\sub{\star}}

\newcommand{\rev}[1]{\textcolor{black}{#1}}

\newcommand{\mhl}[1]{\textcolor{black}{#1}}
\newcommand{\jz}[1]{\textcolor{black}{#1}}
\newcommand{\ja}[1]{\textcolor{black}{#1}}

\title[Ploonets: the fate of detached exomoons]{Ploonets: formation, evolution, and detectability of tidally detached exomoons}

\author[Sucerquia et al.]{\parbox{\textwidth}{ Mario Sucerquia$^{1,2,3},$\thanks{E-mail: mario.sucerquia@udea.edu.co}
Jaime A. Alvarado-Montes$^{1,4,5},$
Jorge I. Zuluaga$^{1},$\\
Nicol\'as Cuello$^{2,3}$ 
and Cristian Giuppone $^{6}.$
}\vspace{0.4cm} \\
$^{1}$ SEAP,  
Instituto de F\'{\i}sica, FCEN, Universidad de Antioquia -
Calle 70 No. 52-21, Medell\'in, Colombia.\\
$^{2}$ Instituto de Astrof\'isica, Pontificia Universidad Cat\'olica de Chile - Santiago, Chile. \\
$^{3}$ N\'ucleo Milenio de Formaci\'on Planetaria (NPF) - Chile.\\
$^{4}$ Centre for Astronomy, Astrophysics, and Astrophotonics, Macquarie University - Sydney, NSW 2109, Australia\\
$^{5}$ Department of Physics \& Astronomy, Macquarie University - Sydney, NSW 2109, Australia.\\
$^{6}$ Universidad Nacional de C\'ordoba, Observatorio Astron\'omico, IATE - Laprida 854, 5000 C\'ordoba, Argentina.\\}

\date{Accepted 2019 July 25. Received 2019 June 26; in original form 2018 November 29\\}

\pubyear{2018}

\begin{document}

\label{firstpage}
\pagerange{\pageref{firstpage}--\pageref{lastpage}}
\maketitle

\begin{abstract}
Close-in giant planets represent the most significant evidence of planetary migration. If large exomoons form around migrating giant planets which are more stable (e.g. those in the Solar System), what happens to these moons after migration is still under intense research. This paper explores the scenario where large regular exomoons escape after tidal-interchange of angular momentum with its parent planet, becoming small planets by themselves. We name this hypothetical type of object a \textit{ploonet}. By performing semi-analytical simulations of tidal interactions between a large moon with a close-in giant, and integrating numerically their orbits for several Myr, we found that in $\sim$50 per cent of the cases a young ploonet may survive ejection from the planetary system, or collision with its parent planet and host star, being in principle detectable. Volatile-rich ploonets are dramatically affected by stellar radiation during both planetocentric and siderocentric orbital evolution, and their radius and mass change significantly due to the sublimation of most of their material during time-scales of hundred of Myr. We estimate the photometric signatures that ploonets may produce if they transit the star during the phase of evaporation, and compare them with noisy lightcurves of known objects (\textit{Kronian} stars and non-periodical dips in dusty lightcurves).  Additionally, the typical transit timing variations (TTV) induced by the interaction of a ploonet with its planet are computed. We find that present and future photometric surveys' capabilities can detect these effects and distinguish them from those produced by other nearby planetary encounters.
\end{abstract}



\section{Introduction}
\label{sec:intro}

\ja{Direct and indirect methods to detect exoplanets} have had \jz{an impressive} success \jz{in the last two decades, as evidenced by the discovery of at least 4102 exoplanets\footnote {For an updated number see \href{www.exoplanet.eu}{www.exoplanet.eu}}} \jz{and a comparably number of candidates}. \jz{Still, most detection techniques are intrinsically biased, resulting in} hundreds of massive exoplanets detected in close-in orbits around their host stars, \jz{the so-called \em Hot-Jupiters}. \ja{But, apart from affecting our assessment of the abundance of planets of all masses located at different distances, hot-Jupiters} provide a great \jz{deal of} opportunities to discover \ja{planetary structures yet unseen beyond the Solar System, e.g. exorings and exomoons}.

\jz{The search of exomoons, for instance, has} led to careful observational \jz{and archive-based surveys, intended to} track over consecutive planetary periods, \ja{signatures of the presence of exomoons using on the one hand traditional techniques (radial velocities and transits) and, on the other hand, more advanced methods like Transit Timing Variation -- TTV, Transit Duration Variation -- TDV or Orbital sampling Effect} (see eg. \citealt{Kipping09,Heller14b,Kipping2016hek}). In this context, moons at large stellar distances (as those in the Solar System) would be hard to detect, while close-in moons could in principle be detectable \ja{(see e.g. \citealt*{Teachey2018})}. \jz{Despite the relentless theoretical and observational efforts, no confirmed exomoon has been detected so far.}

The lack of positive exomoon detections \jz{around close-in gas giants, if not due to our present observational limitations,} could be explained by \ja{diverse planetary mechanisms. For instance,} the perturbations experienced by a moon during its evolution \ja{could drive its disruption or escape}. Thus, in the host-planet migration phase, the moon can either be ejected via ``evection resonance`` \citep*{Spalding2016} or by planet-planet scattering \citep{Gong2013,Hong2018}. As a result, when the moon-planet system reaches \ja{its final configuration in a close-in orbit} \citep{Namouni2010}, the tidal interplay with the planet and the star \jz{would push the moon towards large circumplanetary orbits where it can be perturbed and lost} \citep*{Alvarado2017}. 

\ja{In light of the above, which is a less drastic and fortuitous scenario} as compared to the \jz{previous ones}, moons could potentially be ejected on \jz{relatively} short time-scales. \jz{If some of them survive to the almost unavoidable fate of colliding with the planet or the absorption by the host star}, a \ja{late-type of planetary embryos around the star, or even fully-fledged small planets on their own}, may arise. \jz{This scheme, that has been already considered by several authors \citep{Alvarado2017, Hong2018, Sucerquia2018dm}, will be dubbed hereafter} as the \textit{ploonet scenario}.

\ja{Interestingly, escaped or tidally obliterated exomoons, could play a meaningful role at explaining some of the baffling observational signatures discovered in recent years in the large photometric database provided by \textit{Kepler}, as some of the physical phenomena suggested by those observations seem to be absent in the Solar System}. Among the increasing number of examples, we \jz{may} highlight the puzzling behavior of the lightcurves of KIC-8462852 (Tabby's star, \citealt{Boyajian16}), the spectroscopic evidence of planetary cannibalism in the stars, \jz{the so-called \textit{Kronos} \& \textit{Krios} scenario} (HD 240430 and HD 240429, \citealt{Oh2018}), \jz{or the hypothetical exocometary signatures around} KIC 12557548 and KIC 3542116 \citep{Rappaport2012, Rappaport2018}. \ja{All of these processes, along with their underlying physical mechanisms, still await a proper understanding in the context of planet, moon and ring formation theories}.

The aim of this work is to study the formation and evolution of \jz{hypothetical ploonets} and \jz{their potential connection to some of the previously described puzzling observations.  Moreover, we propose feasible physical mechanisms associated to ploonets that could explain these and other photometric anomalies not observed yet.}

\ja{For that purpose, in \autoref{subsec:ejection} we start by constraining} through semi-analytical calculations, the dynamical star-planet-moon parameters which allow that ploonets \jz{arise in planetary systems}. Then, using these \ja{results} as inputs, we perform N-body simulations in \autoref{subsec:orbital} to obtain the distribution of orbital parameters and survival time-scales. Afterwards, \autoref{subsec:timescales} builds a simple thermodynamical model to calculate the effects of the stellar radiation on the surface and atmosphere of \jz{hypothetical ploonets}. Finally, based on the obtained evolutionary \jz{properties}, we explore in \autoref{subsec:lc} the potential signatures \jz{that these} ploonets \jz{could produce} in lightcurves. To conclude, we discuss and summarize our findings in Section \ref{sec:summary}.

\section{Ploonet formation}
\label{sec:formation}

\subsection{Time-scales of tidal `ejection'}
\label{subsec:ejection}

Early approaches to the problem of orbital tidal-induced evolution of exomoons around close-in planets, assumed that the planet remains unchanged during the moon's orbital motion (e.g. \citealt{Barnes2002}). Recently, \cite{Alvarado2017} proposed a more realistic model for the interaction of planet-moon \jz{systems, including but not restricted to} the evolution of the planetary radius \jz{and the variable mechanical response of the planetary interior to the tides raised by its moons and the host star}. This coupled evolution has proven to be important in star-planet systems too, as shown by \cite{Alvarado2019}.

\jz{According to the model by \citet{Alvarado2017}, which we also apply here}, the most likely fate for exomoons \jz{whose orbit is modified as the result of a tidally-induced interchange of angular momentum with its planet}, is to evolve towards \jz{even further circumplanetary orbits}.  \jz{Time-scales and hence the probability for this to occur during the age of the planetary system, strongly depend on the moon and host-planet initial orbits and other physical parameters}, such as the planet and satellite masses and radii, the initial semi-major axes and the planet core-mass fraction.

For our purposes, we \ja{aim to} identify the subset of parameters that allows the moon to reach \jz{\em large} orbits in \jz{\em short} time-scales, i.e.  times much lower or comparable to planetary system ages, e.g. few Gyrs. \jz{By large orbits we mean orbits comparable or larger than} the critical \textit{secondary Hill's distance} (SHD, see eg. \citealt*{Domingos2006}).  \jz{Beyond this critical distance, the moon orbit becomes unstable}. \jz{Hereafter, we will assume that SHD is located at $0.48 \Rh$ (in the case of prograde satellites) and $0.98 \Rh$ (retrograde satellites) as suggested by the results of \citet{Domingos2006}.  The Hill radius is defined as $\Rh=a_p(M_\star/3M_p)^{1/3}$, with $M_\star$ and $M_p$ the stellar and planetary masses, respectively.} 

\begin{figure*}
    \includegraphics[scale=0.29]{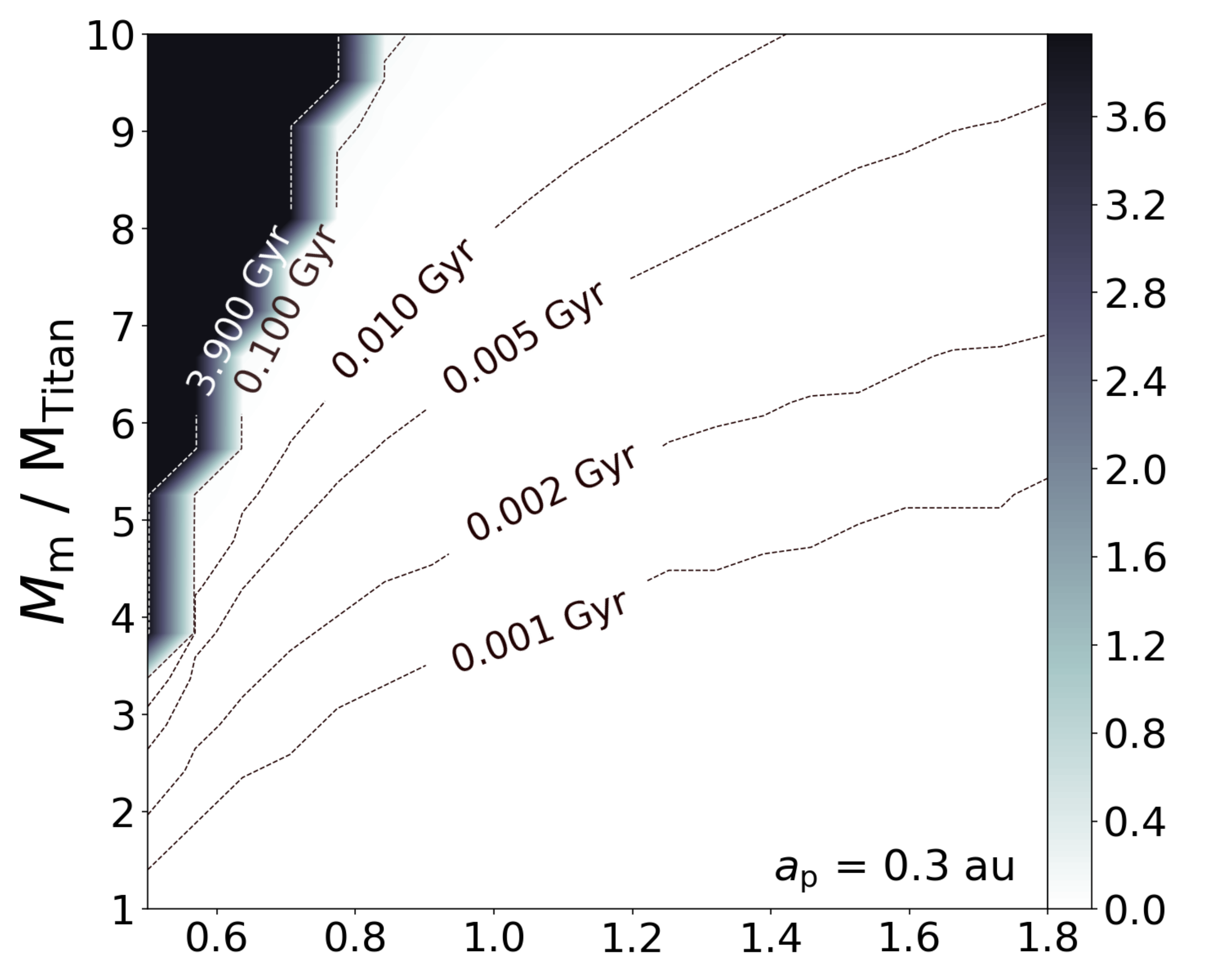}
    \includegraphics[scale=0.29]{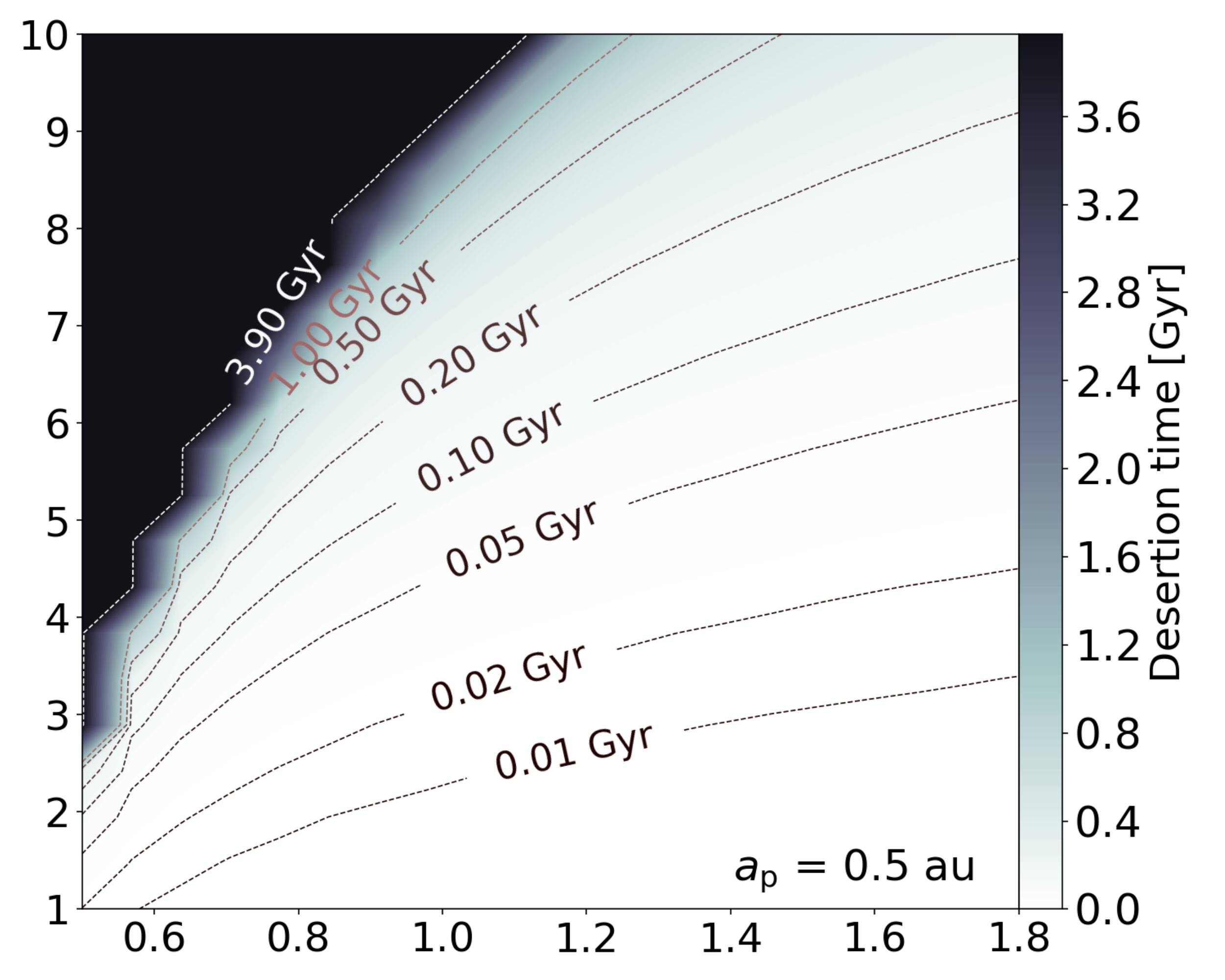}
	\includegraphics[scale=0.29]{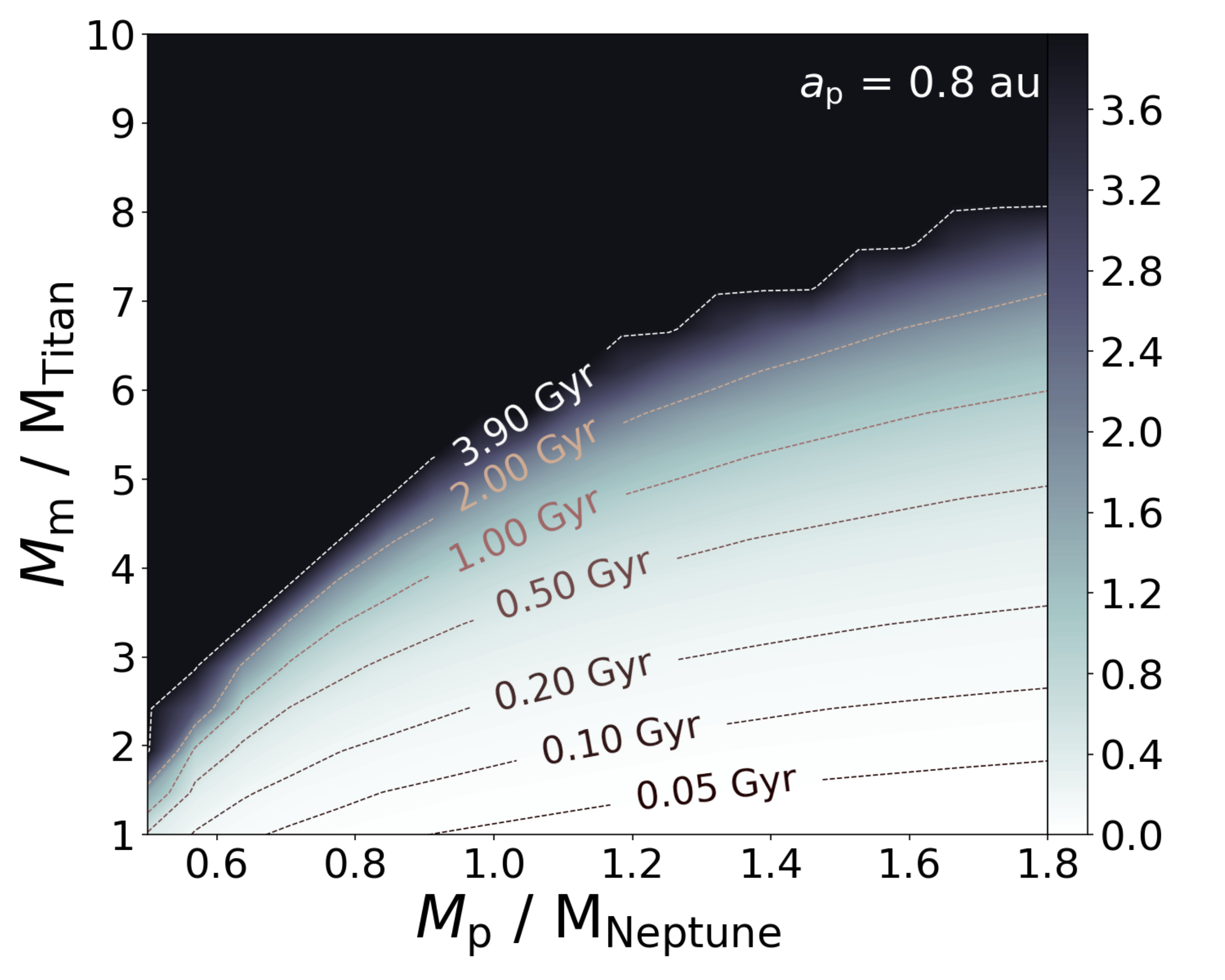}
    \includegraphics[scale=0.29]{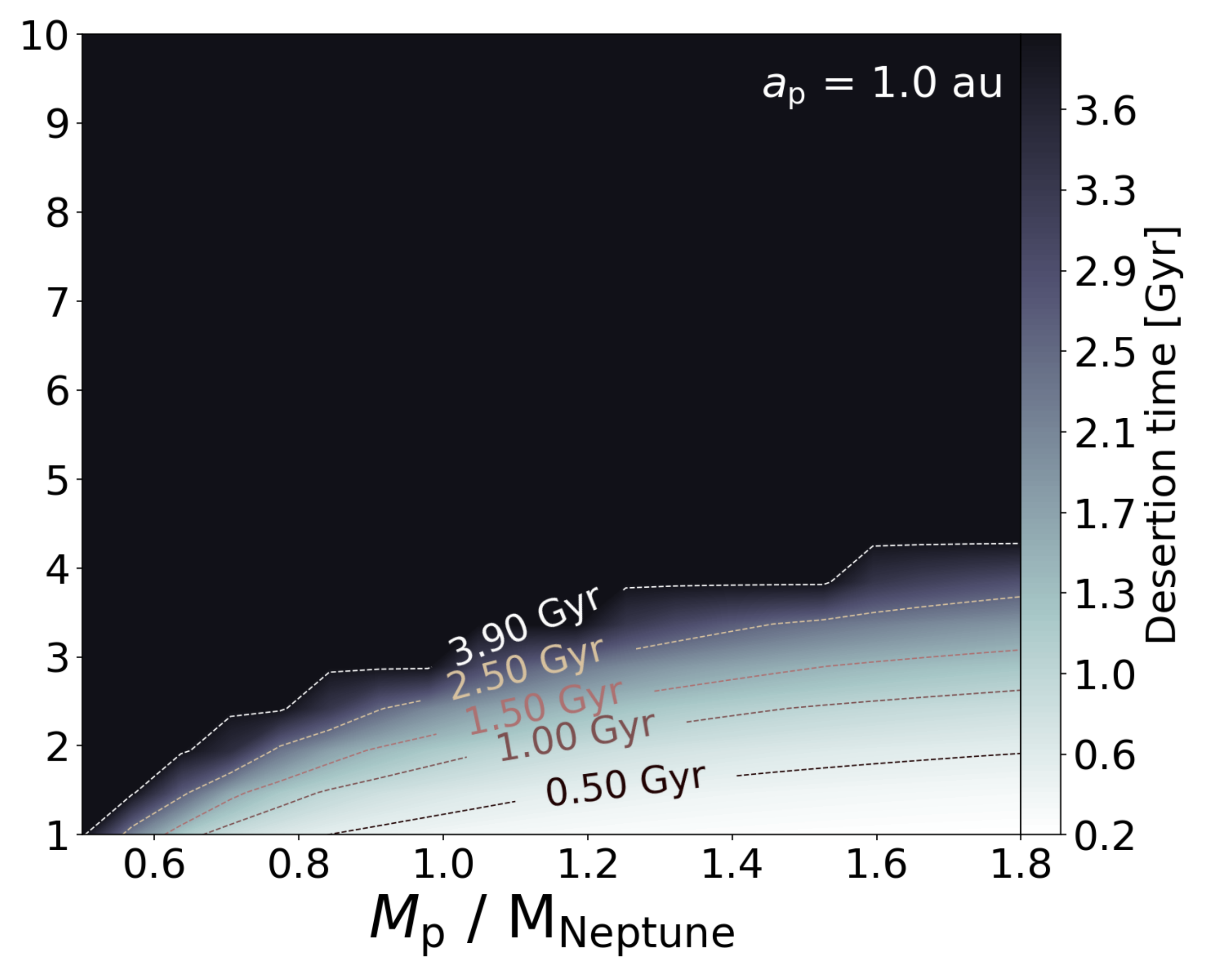}
\caption{Desertion time-scales led by tidal migration for a range of satellite and planet masses, and regarding to each planetary semi-major axis $\ap$. }
\label{fig:tidev}
\end{figure*}

The left-hand panel of Fig. \ref{fig:tidev} presents the outcome of a set of semi-analytical calculations of the {\em desertion time}, \jz{namely, the time required by a moon to reach the SHD}. In these simulations, the planet and satellite masses range between 0.5 to 1.8 Neptune masses and between 1 to 10 Titan masses, respectively. We assume a solar-mass star and a planet in a fixed circular orbit with a=0.3, 0.5, 0.75 and 1.0 au. 
\mhl{Moreover, we fixed in our simulations the initial $\am$ at $3.0 \;\Rp$ and the planetary rotational period at $T_p=15.7$ hours. For the planet interior we set $\alpha$, $\beta$ and $\gamma$  defined as:}
\begin{eqnarray}
\alpha = \frac{R_\mathrm{c}}{R_\mathrm{p}} = 0.249, \;\;\; \beta = \frac{M_\mathrm{c}}{M_\mathrm{p}} = 0.112, \;\;\; \gamma = \frac{\alpha^3 \, (1-\beta) }{\beta \, (1-\alpha^3)}\; ,
\end{eqnarray}

\mhl{which are dimensionless parameters defined in terms of the bulk properties of the planet $R_\mathrm{c}$, $R_\mathrm{p}$, $M_\mathrm{c}$, $M_\mathrm{p}$ which are the core radius, planetary radius, core mass and planetary mass, respectively. $\gamma = \rho_\mathrm{e} / \rho_\mathrm{c}$ is defined as the ratio between the planetary core and (liquid) envelope density  (see equation 2 in \citealt{Alvarado2017}).}

The darker area in Fig.  \ref{fig:tidev} represents the subset of parameters for which the moon desertion times are greater than 4.0 Gyr. In our simulations, $\sim20$ per cent of moons randomly located around a planet at $\ap=0.5$ au, desert in times larger than this limit. On the other hand, the lighter area in the plot represents the subset of unstable moons that are ejected and can potentially become ploonets ($\sim80$ per cent of the moons in our simulation). The fraction of unstable moons strongly depends on the planetary semi-major axis, $\ap$. For others values of $\ap$, we find that the fraction of {\em potential ploonets progenitors}, can vary from 90 to 20 per cent.

\rev{In a planet-moon system we have that \citep{Murray2000}, 
\beq{eq:nmoon}\frac{\Der n_\mathrm{m}}{\Der t}\propto -\frac{k_{2, \mathrm{p}}}{Q_\mathrm{p}}\frac{M_\mathrm{m}}{M_\mathrm{p}^{8/3}}\hspace{0.5cm}\text{and}\hspace{0.5cm}\frac{\Der \Omega_\mathrm{p}}{\Der t}\propto -\frac{k_{2, \mathrm{p}}}{Q_\mathrm{p}}\frac{M_\mathrm{m}^2}{M_\mathrm{p}^3}\;.\eeq}
\rev{where $k_{2, \mathrm{p}}$ is the second-order love number, and $Q_\mathrm{p}$ the tidal quality factor.}

\rev{The planet-moon system evolution is driven by the spin-orbit angular momentum exchange, and the tidal-dissipated energy per rotational period from the planet makes the moon to decrease its initial mean orbital motion $n_\mathrm{m}$. At the same time, the planet's rotation $\Omega_\mathrm{p}$ decreases due mainly to the stellar torque, and eventually  $\Omega_\mathrm{p}\sim n_\mathrm{m}$ at the so-called synchronous radius (or critical semimajor axis as defined by \citealt{Barnes2002}).} 

\rev{We must emphasize that the `desertion time-scales' in Fig. \ref{fig:tidev} are based on the calculations of \cite{Alvarado2017} in the `realistic scenario', where planetary tides are evolving over time along with the orbital evolution of the moon. Within this framework we have that the planet's tidal-dissipated energy is directly proportional to $\Omega_\mathrm{p}$, i.e. $k_{2, \mathrm{p}}/Q_\mathrm{p}\propto\Omega_\mathrm{p}^2$ \citep{Alvarado2017}, which in turn decreases the planet's dissipated energy while the rotational braking occurs. The most important effect of these two coupled processes, namely the evolution of the planetary dissipation parameters and the orbital evolution of the moon, is that once the moon arrives to the synchronous radius it is stalled there and further outwards migration takes place at a very slow rate. However, the moon does not return because the planet's rotation spins down too slowly, so the torques are not enough to pull the moon back to closer distances.}

\rev{From equation (\ref{eq:nmoon}) and Fig. \ref{fig:tidev} we can see that larger moons migrate outwards slowly, but make the planet to decrease its rotation at an even slower rate. Therefore, the synchronization takes longer (dark areas in Fig. \ref{fig:tidev}). Also, for a fixed moon mass it is noticeable that larger planets make the moon migrate outwards at a slower rate than the spin deceleration of the planet, so $\Omega_\mathrm{p}$ can reach $n_\mathrm{m}$ in shorter times (light areas in Fig. \ref{fig:tidev}). } 

\begin{figure*}
    \includegraphics[scale=0.59]{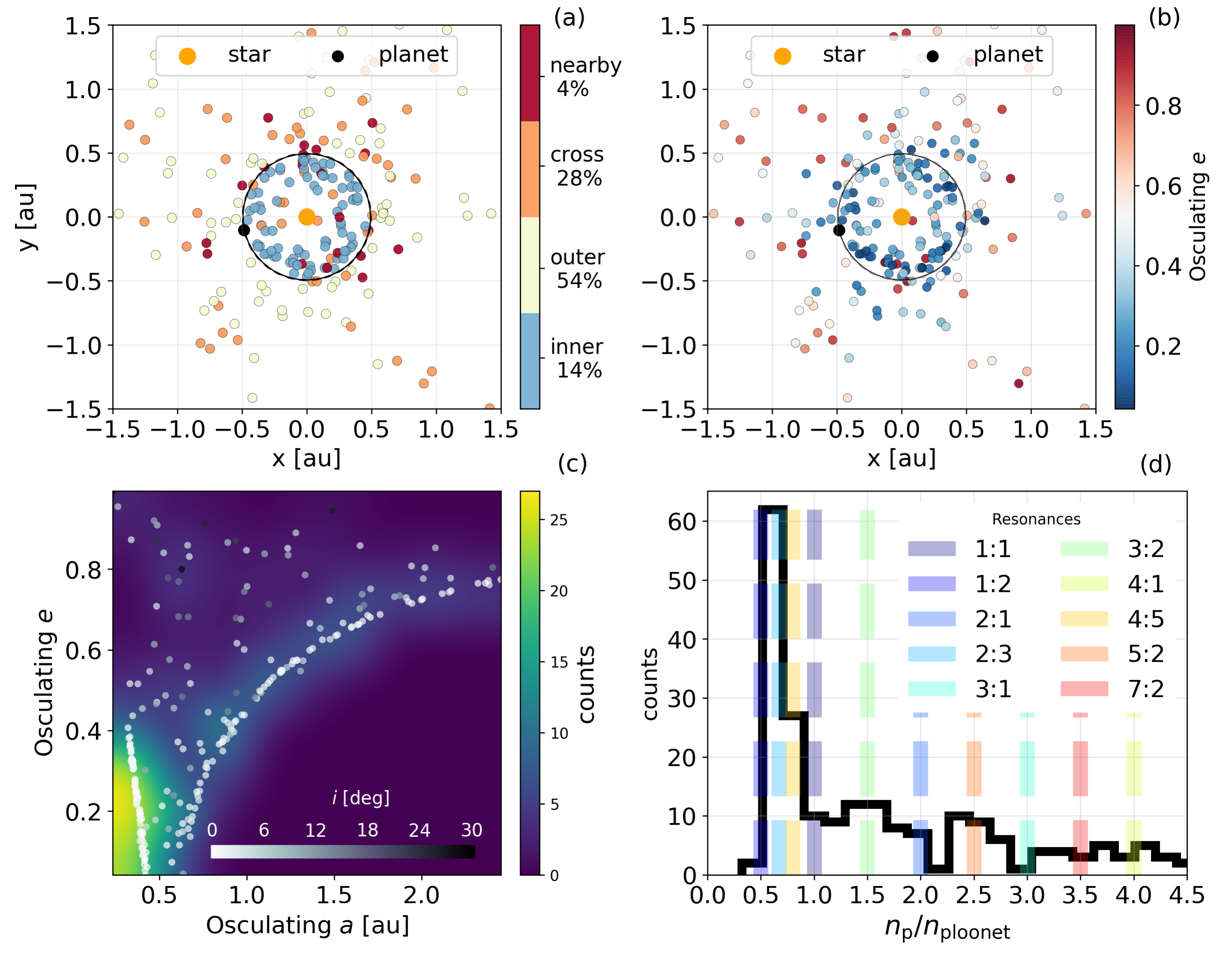}
\caption{(a) and (b) orbital characterization for \textit{ploonets}, in colors, the type of orbit and the ploonet's eccentricities after 0.5 Myr. (c) Orbital parameters distribution. $\sim 48$ percent of the initial amount of particles survived and contribute to the statistics. (d) Frequency histogram of the occurrence of instantaneous orbital resonances in the system.}
\label{fig:nbody}
\end{figure*}

\subsection{\jz{Circumstellar} orbital evolution}
\label{subsec:orbital}

Once a moon reaches an unstable orbit around the planet, \jz{it may escape from the Hill-sphere and reach a circumstellar orbit}. In order to compute the distribution of ejected moon orbital parameters and its evolution, \jz{we perform Monte Carlo} N-body \jz{simulations}. We start by throwing 1\,000 test particles placed \jz{initially} at a \jz{circular orbit} with $\am=0.48$ $\Rh$ with a small inclination (10$^\circ$) and random \jz{initial circumplanetary} true anomaly $f_\mathrm{m}$. \mhl{We selected this small inclination because after several experiments (zero and small inclination measured respect to the orbital plane of the planet), we note that the chances of an ejected moon for surviving after numerical integration is slightly larger than in the case of coplanar orbits.} 

 \jz{We integrate the three-body system, star, planet and moon,} for a timespan of 0.5 Myr. \ja{This integration time, although not enough to} guarantee the long-term orbital stability \jz{of a ploonet around the star}, \jz{it is large enough} to \jz{study} the effects of this type of objects on young planetary systems (see Section \ref{sec:detectability}).  All of the integrations were performed using the well-tested code \href{https://janus.astro.umd.edu/HNBody}{\tt HNBody} \citep{Rauch2012}. Moreover we verify the outcomes of our simulations using {\tt REBOUND} \citep{Rein2012}, obtaining similar results.

Fig. \ref{fig:nbody} shows the outcome of our N-Body simulations in the case of \jz{a planet (having a mass $m_\mathrm{p}=10^{-4}$ the mass of the star) in a circular orbit about a solar--mass star, with $\ap=0.5$ au}. 

We find that $\sim$54 per cent of the ploonets, ends-up in orbits larger than the planetary distance (labeled \textit{outer} ploonets) and $\sim$14 per cent of them ends-up inside (labeled \textit{inner} ploonets). Almost one third ($\sim$28 per cent) of them after leaving the Hill's sphere, ends-up in highly eccentric orbits that crosses the planetary orbit (labeled \textit{crossing} ploonets). Finally $\sim$ 4 per cent of the objects after the integration have orbits very similar to that of the planet (i.e $a=a_\mathrm{p}\pm 0.1$).  We call these objects \textit{nearby} ploonets. On the other hand, \jz{although there is a large diversity of final eccentricities (panel b in Fig. \ref{fig:nbody})}, a significant fraction of \ja{inner ploonets} with low eccentricity orbits and laying close to the planet were ignored in our experiment. 

Something important to highlight is that ploonets colliding with the planet ($\sim$44 per cent) or being absorbed by the star ($\sim 6$ per cent), as well as those ejected from the planetary system ($\sim2$ per cent), were pruned out and not accounted for in the statistics. The ploonets represented in  Fig. \ref{fig:nbody} correspond to $\approx 48$ per cent of the total number of initial simulated objects, that survived during the integration time.

In panels (c) and (d) of Fig. \ref{fig:nbody}, the distribution of the final osculant orbital elements, $a_\mathrm{s}$, $e_\mathrm{s}$ and $i_\mathrm{s}$ \jz{of the ploonets with respect to the star, are shown}. We observe in a peaked distribution of low-eccentric orbits \jz{with similar size than that of the planet (nearby ploonets)}. \jz{As expected, non nearby ploonets are distributed in the osculating $a_\mathrm{s}-e_\mathrm{s}$ plane following the characteristic pattern of constant Tisserand parameter}. The periapsis argument is \jz{almost uniformly distributed} (not shown in Fig. \ref{fig:nbody}) and thereby no preferred \jz{orbital} orientation is found. 

Finally, the distribution of the mean orbital motion of the surviving exomoons is plotted in panel (d) of Fig.  \ref{fig:nbody}.  There are three recognizable peaks close or at resonances (planet:moon) 1:2 (inner ploonets), 3:2 and 5:2 (outer ploonets). It is worthy to mention that the resonances found are transients states of the system. Moreover, almost all the orbits resulted chaotic/unstable according to the assessing of the Mean Exponential Growth factor of Nearby Orbits (MEGNO) parameter \citep*{Cincotta2003}, according to the implementation provided by {\tt REBOUND} \citep{Rein2015}. This calculation is not shown in the plot.  

\section{Ploonet detectability}
\label{sec:detectability}

\jz{If according to the previous analytical and numerical models, a population of small ploonets could arise in young planetary systems hosting close-in giant planets, and we additionally assume that some of them survive to the long-term orbital evolution of the system, the next question we should address is what are the observable signatures that evidence the presence of these objects?.}

Besides the direct photometric or indirect stellar radial-velocity detection (that could be hard to distinguish from the observational signature of an actual small planet), we consider and evaluate here, two additional and distinctive observational signatures of ploonets: 1) the transit of rapidly evaporating planetary embryo-sized objects with an orbit closely related to an already detected close-in giant planet and 2) characteristic TTVs of the parent planet transit signal.


\jz{The very existence of ploonets depends on two basic conditions: 1) planetary moons around giant close-in planets are formed before or during the migration process when abundant circumstellar gas and dust is accreted by the planet \citep{Namouni2010,Heller2015a}; and 2) the moons survive planetary migration \citep{Namouni2010,Spalding2016}.}  \jz{Although it is hard to ensure that both conditions will be fulfilled under general circumstances, the characteristic time-scales of moon formation ($10^3-10^5$ years, \citealt{Canup2006}) and planetary migration ($10^3-10^7$ yrs \citealt{Armitage2010}) seem to favor the idea that some migrating giant planets could arrive to their final orbit, carrying along several volatile rich moons.}

\jz{So far, we have studied the tidal and orbital evolution of those objects.  Now it is worth asking what would happen to their rich volatile content while exposed to the high radiation levels of the final planetary orbit.} \mhl{This is a worthy question to address because icy bodies lying at close distances to their star, can undergo strong enough surface and atmospheric processes to be noticeable in lightcurves. In contrast, moons still in circumplanetary orbits could be still protected by the magnetic field of their planets  \citep{HellerZ2013}, reducing the rate of atmospheric erosion produced by the stellar wind.}

\jz{Three intertwined physical processes affect an icy-rich object subject to high levels of stellar radiation}: 1) the evaporation or sublimation of its icy surface; 2) \jz{the arising of a massive, optically thin transient atmosphere whose mass and volume rapidly evolve in time}; and 3) the erosion and eventual loss of its envelope which leaves a trace of gaseous and dusty material along its orbit. 

\jz{How fast are those processes?. If most of the involved process are very fast, the probability of detecting their observational signatures will be rather small.  If, on the other hand, the timescales are large enough, we could have real chances to detect them in the near and middle future.  In the following section we model the evaporation of a pure-ice ploonet subject to large stellar insolation.}

\subsection{Ploonets' ``evaporation'' time-scales}
\label{subsec:timescales}

\jz{Computing the time-scales for the ``evaporation'' of a close-in ploonet, namely, the loss of a significant amount of volatile and refractory material while exposed to high levels of radiation vicinity of its host star, especially in the case of a complex planetary object, is challenging.  Real moons are composed of a normally differentiated admixture of volatile and refractory materials, which are differently affected by radiation and temperature.  Their thermodynamic properties will change in a complex manner as large amount of volatiles are loss (see e.g. \citealt*{Lehmer2017}).  In a realistic situation, even under the harshest radiation conditions, a ploonet will not entirely evaporate, but probably leave a residue of a refractory material.}

\jz{In order to get a first order estimation of the typical timescales and rates we will model the evaporation of a pure water ice ploonet.  As expected, the timescales computed in this case will be an underestimation of the actual timescales for a more complex object, as argued below.}

\jz{The equilibrium temperature of an airless, rapidly rotating body (rotation period much shorter than the orbital period) having a bond albedo $A$, orbiting a star with a constant luminosity $\Lstar$ in a circular orbit of radius $\aploon$, is given by}: 
\begin{equation} 
\Teq=\left[\frac{\Lstar \, (1-A)}{16\,\pi\,\SigmaSB \,\aploon^{2}}\right]^{1/4},
\end{equation}

\noindent
\rev{with $\SigmaSB$, the Stefan-Boltzmann constant.  If a low-pressure gaseous envelope is created as the ice sublimates, the surface temperature of the ploonet will be slightly larger, which can be estimated by computing the effect of IR absorption of water (greenhouse effect) under the gray, radiative, plane-parallel approximation:}
\begin{equation} 
\rev{ T=\,\left[\left(1+\frac{\tau}{2}\right)\frac{\Lstar\,(1-A)}{16\,\pi\,\SigmaSB \, \aploon^{2}}\right]^{1/4},}
\end{equation}

\noindent \rev{where $\tau$, the total gray atmospheric optical depth in the thermal IR at the
surface, is given by} 
\begin{equation}
   \rev{\tau = \frac{\kappa_\mathrm{ref}} {2}\, \left( \frac{P}{P_\mathrm{ref} } \right)^\alpha,    }
\end{equation}

\noindent \rev{ with the mass absorption coefficient, $\kappa_\mathrm{ref}$, equal to 0.05 m$^2$ kg$^{-1}$, at a reference pressure $P_\mathrm{ref} = 10^{-4}$ Pa, and surface pressure $P$. Here,  $\alpha=1$ or $2$, according to the pressure regime (see e.g \citealt{Lehmer2017} and references therein.)}

\jz{Assuming that evaporation happens in a low pressure atmosphere \rev{(i.e $\alpha=1$)}, the ice mass-loss rate from the surface, $\dot{M}_\mathrm{s}$ can be constrained by the rate $Z_r$ (measured in kg m$^{-2}$ s$^{-1}$) at which water ice sublimates at temperature $T$ from the surface of a an object with radius $\rploon$ \citep{Estermann1955,Bohren1998}}: 

\begin{equation}
Z_{r} = e_\mathrm{sat}(T)\left(\frac{m_\mathrm{w}}{2\pi RT}\right)^{1/2}\exp\left[\frac{2\, m_{w}\,\sigma_{i}\,(T)}{\rho_{i}(T)\, \rploon \,RT}\right] \,\,\,.
\end{equation}
\jz{Here, $m_\mathrm{w}$ is the molecular weight of water and $R$ the universal constant of ideal gas and $e_\mathrm{sat}(T)$, $\sigma_\mathrm{i}(T)$, $\rho_\mathrm{i}(T)$ are the saturation vapor pressure, the surface ice tension and  the surface ice density at temperature $T$. For the latter quantities we use standard thermodynamic expressions from \citet{Andreas2007}}.  

\jz{The total water mass loss rate from the ploonet surface is}:

\begin{equation} 
\label{eq:Msurf-rate}
\dot{M}_\mathrm{s} = - Z_{r}S\sub{pl} = -Z_{r}\left[ \frac{6\,\sqrt\pi\,\mploon}{\rho_\mathrm{i}(T)}\right]^{2/3},
\end{equation}
\noindent
\jz{where the ploonet surface area $S\sub{pl}$ has been expressed in terms of its mass $\mploon$ and average density which is assumed constant all across the object and equal to the surface ice density $\rho\sub{i}(T)$}.

\begin{figure}
{ 
\centering 
\includegraphics[scale=0.42]{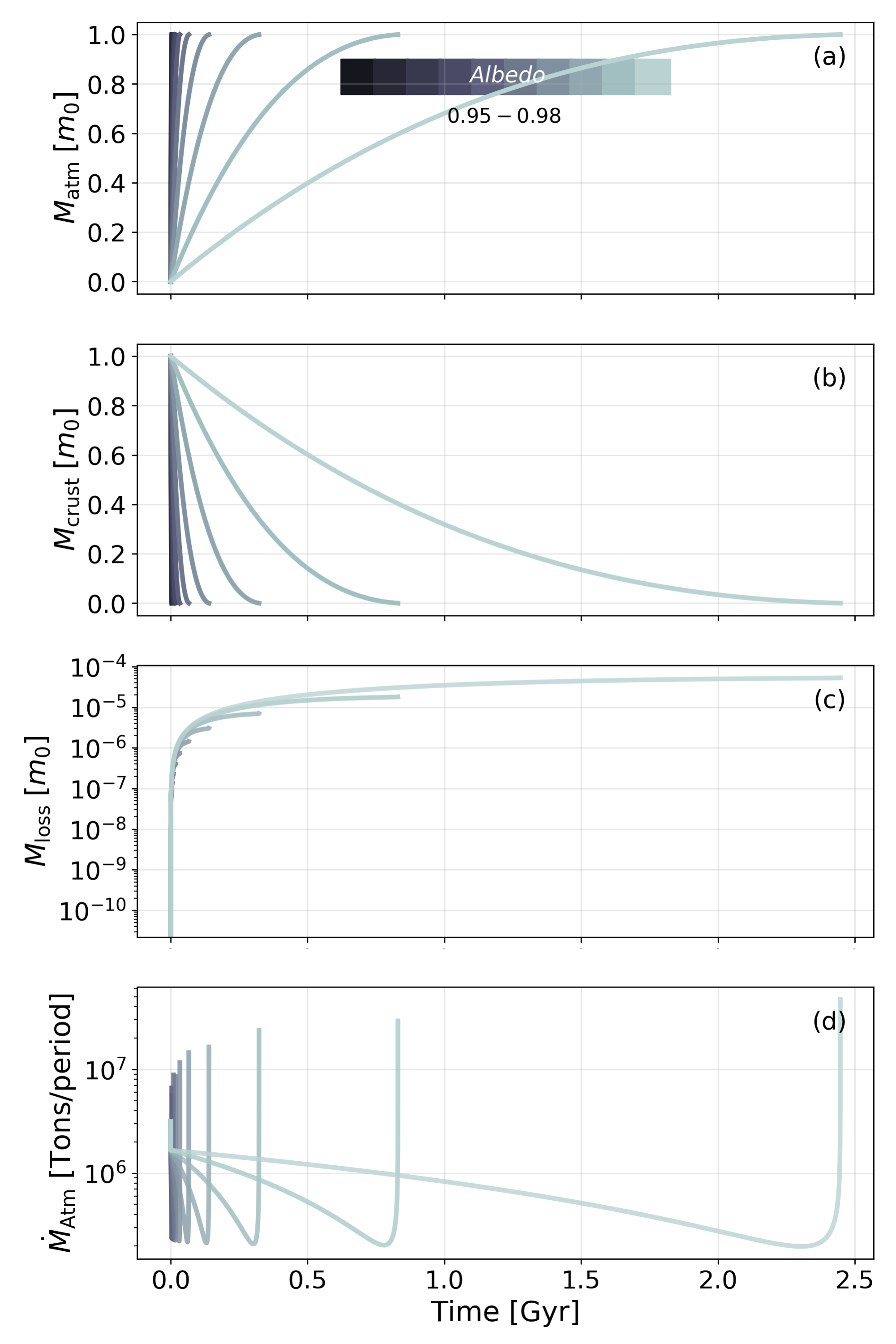}
\caption{
\rev{(a) The process of atmospheric growth ($M_\mathrm{atm}$) in terms of the initial mass of the ploonet ($m_\mathrm{0} \sim 10^{23}$ kg). (b) surface sublimation, (c) the integrated mass loss, and (d) the mass-loss rate evolution of a ploonet is shown for different albedos ranging from  0.95 to 0.98 (colorbar)}.
}
\label{fig:mass-evol}
}
\end{figure}

\begin{figure}
{ 
\centering 
\includegraphics[scale=0.34]{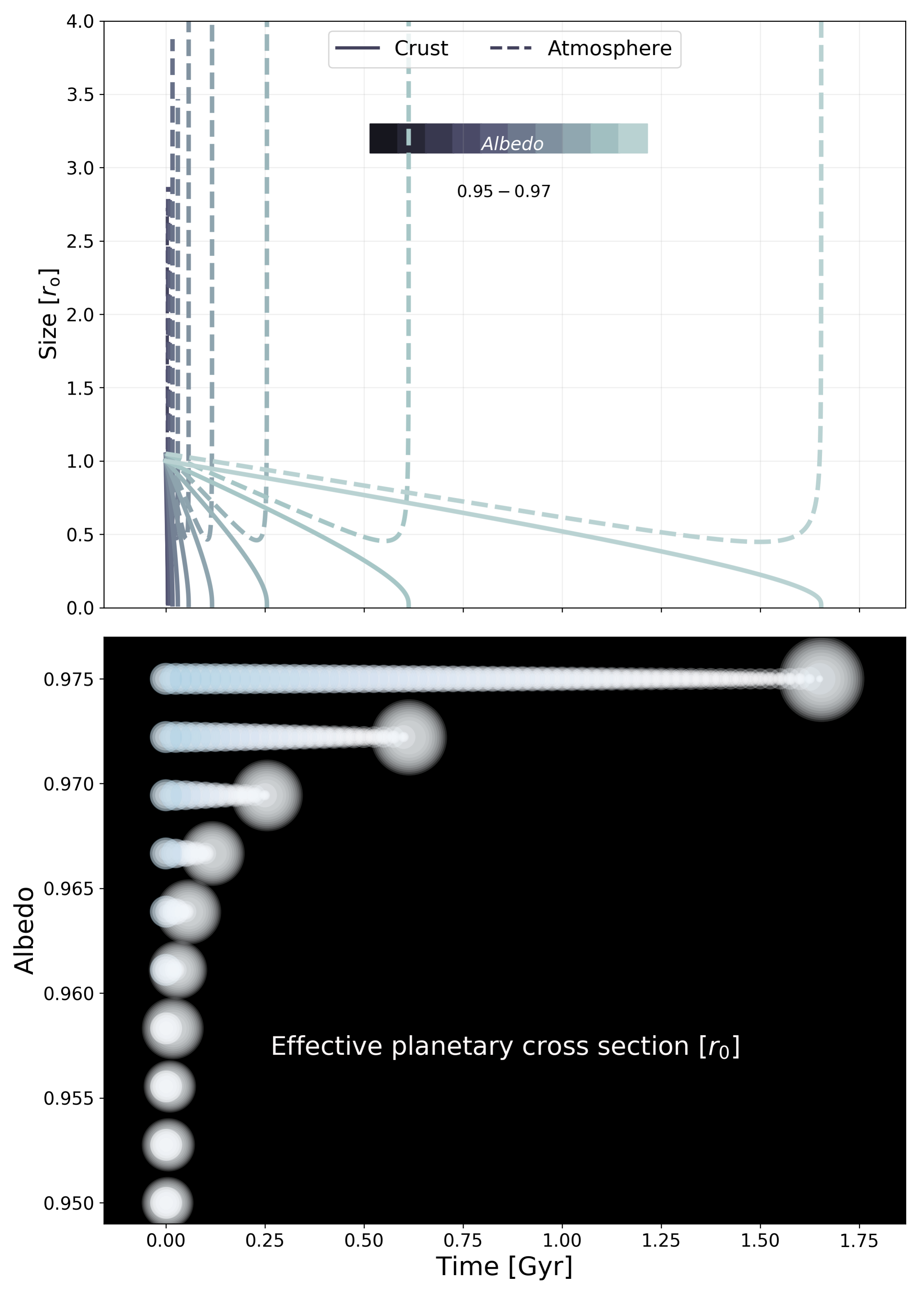}
\caption{
\rev{{\it Top panel}: The variation of the crust (solid lines) and atmosphere sizes (dashed lines) over time for a range of surface albedos. {\it Bottom panel}: evolution of the effective cross section ($h+r\sub{pl}$) of a ploonet with different albedos. For example, the bond albedos of Eris (a dwarf planet) and Enceladus (a Kronian moon) are about 0.96 and 0.99, respectively.}
}
\label{fig:radius-evol}
}
\end{figure}

\jz{Once sublimated, the transient water rich atmosphere of our body is subject to two erosion effects: 1) XUV-induced atmospheric escape \citep{Sanz-Forcada2011} and 2) stellar-wind drag \citep*{Zendejas2010}.  Although both effects can be largely different in magnitude and are mostly independent, we will constraint the total mass-loss as a simple linear combination of both:}

\begin{equation}
\label{eq:Mloss-rate}
\dot{M}_\mathrm{loss} = -\frac{3F_\mathrm{XUV}}{4G\rho} \; - \; \left(\frac{r\sub{pl}}{a\sub{pl}}\right)^2 \frac{\dot M_\star \alpha}{2} \,\,\,,
\end{equation}

\noindent
\jz{where $F_\mathrm{XUV}$ is the XUV (X-rays and extreme Ultra Violet)  stellar flux, $M_\star$ is the stellar mass loss rate, $\alpha=0.3$ is the entrainment efficiency, and $G$ the gravitational constant. All these quantities will be estimated with the models introduced and used in \citet*{Zuluaga2016}}.

\jz{Using the sublimation rate at the surface (equation \ref{eq:Msurf-rate}) and the gaseous envelop mass-loss rate (equation \ref{eq:Mloss-rate}), the atmospheric mass $M\sub{atm}(t)$ will simply obey the continuity equation:}

\begin{equation}
\label{eq:Matm-rate}
\dot{M}_\mathrm{atm} = \dot{M}_\mathrm{loss} - \dot{M}_\mathrm{s}
\end{equation}

\jz{As the atmospheric mass evolve, the pressure $P$ and scale-height $h$ of the atmosphere change. A first order approximation of these quantities, assuming a homogeneous atmosphere in hydrostatic equilibrium in an uniform gravitational field, are given by } 

\begin{equation}
\label{eq:P}
\rev{ P(t) \,= \, \frac{n\, R\,T }{ V\sub{atm}} \,=\, \frac{M\sub{atm}\,R\, T }{g_\mathrm{pl}\, m_\mathrm{w}\, V\sub{atm}}}
\end{equation}

\noindent
and

\begin{equation}
\label{eq:h}
\rev{h\,(t)= \frac{R\, T}{g_\mathrm{pl}\, m_\mathrm{w}},}
\end{equation}

\noindent
\rev{ where $n$ is number of moles of the atmospheric water vapour, $V\sub{atm} = (r\sub{pl}+h)^3 - r\sub{pl}^3$ the volume of its gaseous envelope, $T$ the instantaneous temperature, and  $g_\mathrm{pl}$ the ploonet's gravitational field intensity.} 

\jz{Plugging-in all these effects on the mass-loss rate equations (\ref{eq:Msurf-rate}, \ref{eq:Mloss-rate}, and \ref{eq:Matm-rate}), we can predict the evolution of the mass and radius of the ploonet and from there constraint the timescale of evaporation of its volatile content.}

\rev{In Fig. \ref{fig:mass-evol} we present the evolution of the mass of a ploonet located in a 0.5-au circular orbit around a solar-mass star, as computed with our simplified phenomenological model. Different values for the bond albedo are assumed. In (a), (b), and (c) the masses (y-label) are given in terms of the initial mass of the ploonet $m_\mathrm{0}$, which is equal to  $m_\mathrm{0} \sim 10^{23}$ kg.} The evaporation time-scale is strongly dependent on the albedo, and increasing slightly the albedo makes the evaporation time-scales larger. In all cases, however, evaporation time-scales are of the order of hundred of Myrs, and can even reach several Gyrs. Since the time-scales of our phenomenological model are likely an underestimation of the actual time-scales, these results imply that the probability of observing a ploonet during these transient processes is relatively high.

In Fig. \ref{fig:radius-evol} the evolution of the effective radius (i.e the radius of the remaining solid mass and that of the gaseous envelop) of a ploonet is presented (upper panel). To better illustrate the evolution in the size of the icy ploonet, we have additionally depicted (lower panel) the effective cross section for three bodies with a bond albedo $A$ ranging from 0.86 to 0.9. In our Solar System, bond albedo ranges between $\sim 0.09$ for Mercury and $\sim 0.96$ for Eris, which is a dwarf planet.

\begin{figure}
{ 
\centering 
\includegraphics[scale=0.40]{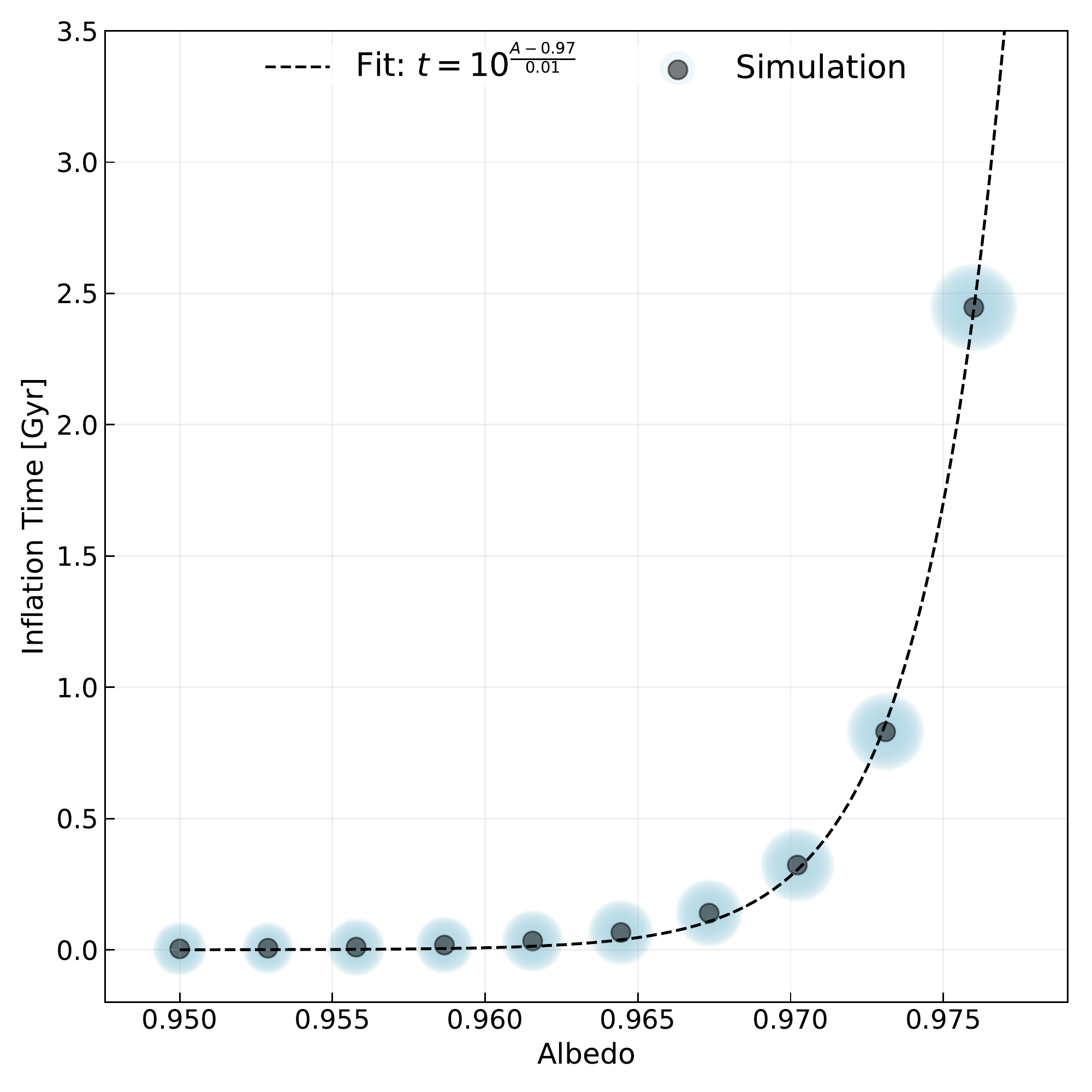}
\caption{\rev{Power-law fit of the inflation time of ploonets as a function of the bond albedos. The dots represent the simulation data for different ploonets, and are surrounded by a spherical shadow to depict the final size reached by a ploonet with a given albedo.}}
\label{fig:fit}
}
\end{figure}

\rev{It is worth noting that surfaces with lower values of $A$ are quickly sublimated, which in turn creates a large gaseous envelope that decelerates the sublimation rate of their crust. In all the cases studied here, a ploonet could multiply its original apparent cross-section after several hundreds to thousands of Myr. That said, a power-law fit can be performed (as a function of their bond albedo), to find the inflation time-scale at which the cross-section of the ploonets \textit{inflates} until reaching the maximum. The fitting procedure gives $t=10^{(A-0.97)/0.01}$ (see Fig. \ref{fig:fit}).}

Our results rely on a very simplistic phenomenological model.  We do not model the ice-melting processes (which for instance may change the Albedo) or the build up of a substantial atmosphere which may affect the greenhouse effect or other second order processes probably involved in the volatile content evolution of the ploonet \citep{Lehmer2017}. Nevertheless, we are confident that our model is able constraint the time-scales of the involved processes at least for the purpose of testing the hypothesis of the ploonet signature detection.


\subsection{Ploonet's observational signatures}
\label{subsec:lc}

Ploonets are by definition small objects (they are former satellites). In spite of that, they could generate notorious imprints in the flux of their parent stars via the generation of noisy transits associated to their \textit{evaporation} process (\autoref{subsec:timescales}).  These imprints may eventually resemble those observed around  KIC 12557548 and KIC 3542116 \citep{Rappaport2012,Rappaport2018}, and others. Even if they are undetectable via transits (if, for instance, their orbits are tilted), they could still affect the periodicity of the transits of their progenitor planet.

\begin{figure}
{ 
\centering 
\includegraphics[scale=0.33]{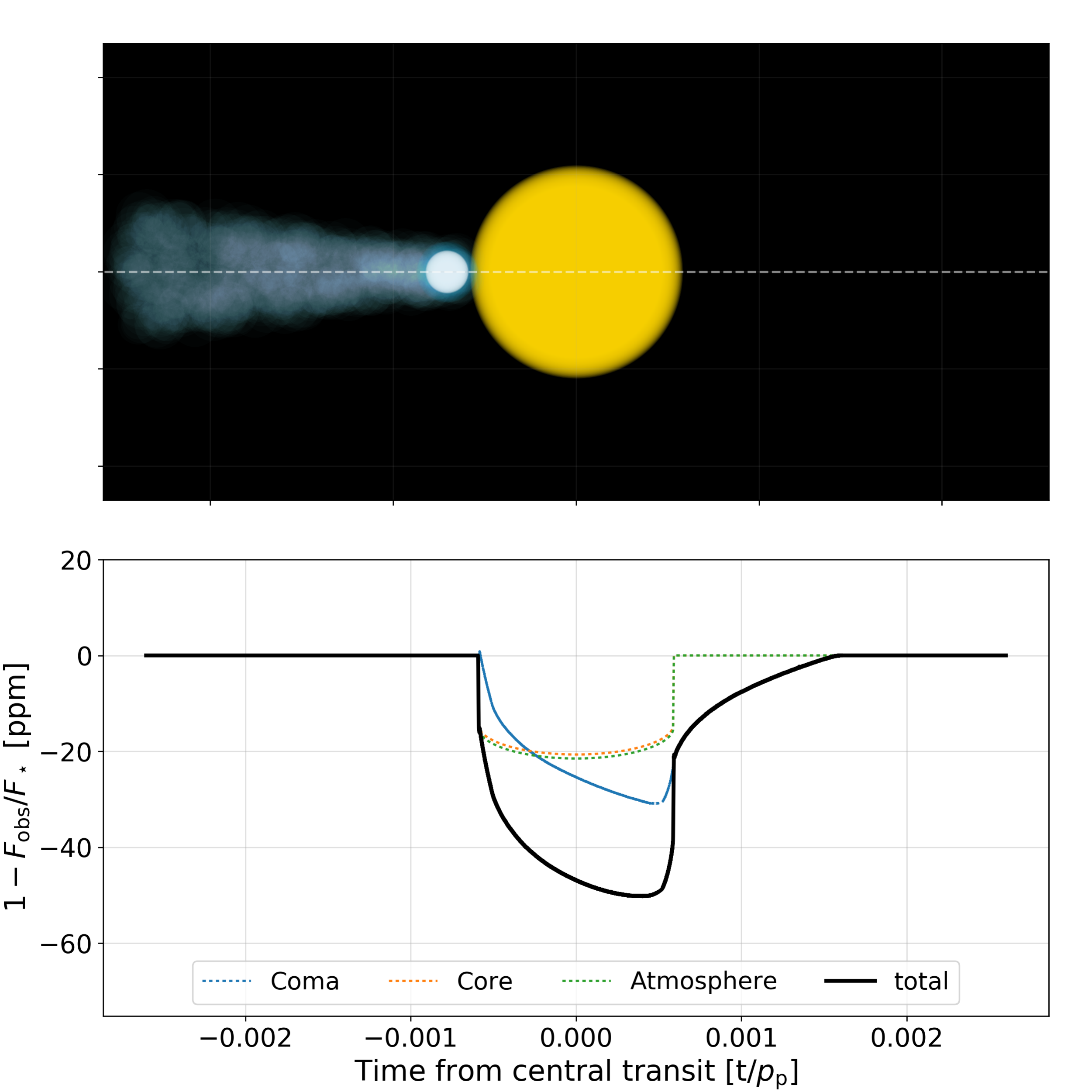}
\caption{Expected signature of ploonets in transit signals. The core and the atmosphere are modeled as circles, The dusty tail is composed of thousand of small particles, traced by adopting a power law of slope $-1.5$. $F_\mathrm{obs}/F_\star$ is the normalised observed stellar flux.  }
\label{fig:transit}
}
\end{figure}

\begin{figure*}
{ 
\centering 
\includegraphics[scale=0.5]{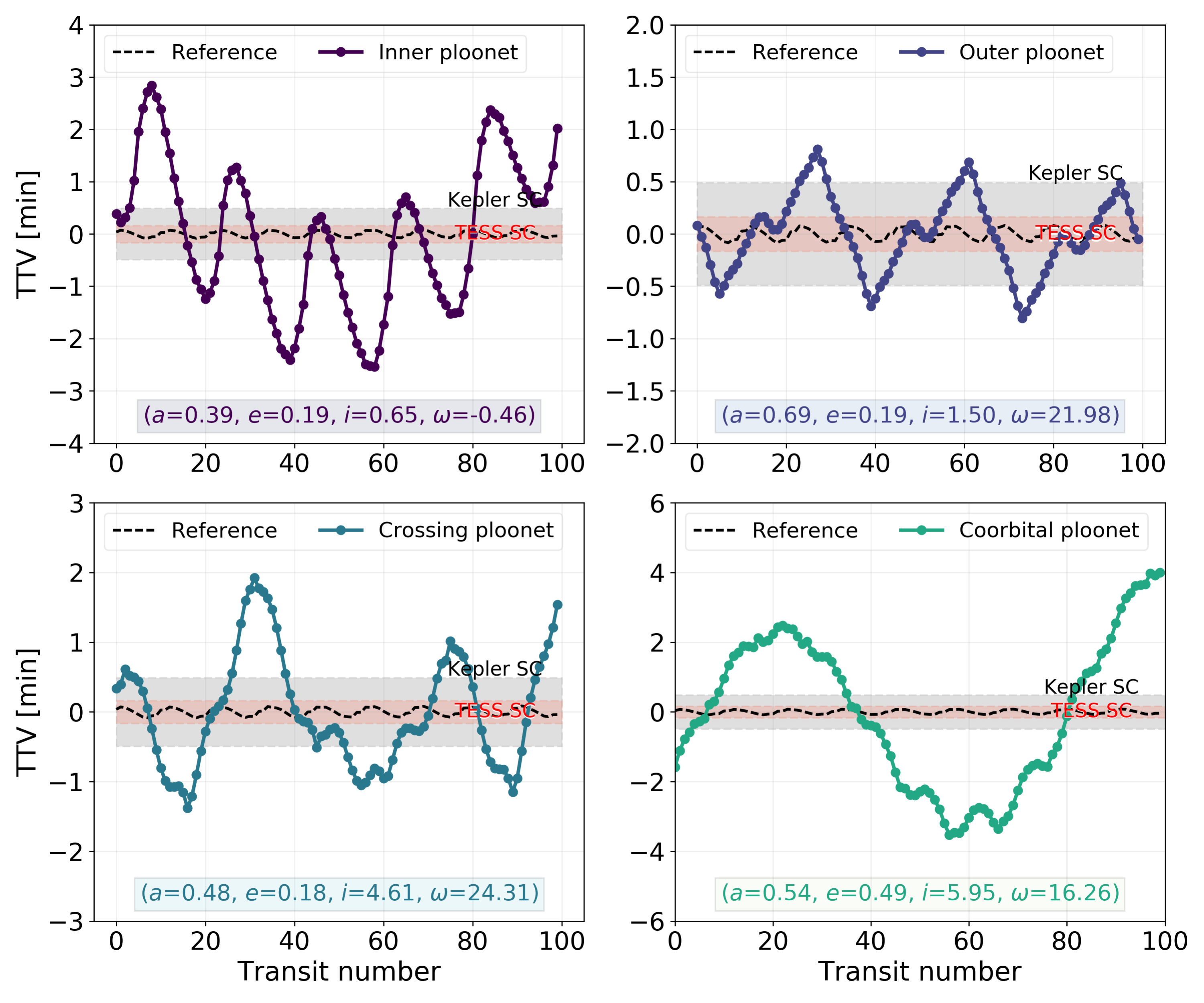}
\caption{Transit Timing Variations for the parent planet due to the disturbing effect of a ploonet regarding to their orbital classification (\textit{inner, outer, crossing} and \textit{nearby}) mentioned in \autoref{subsec:orbital}. Kepler SC and TESS SC (gray and red strips, respectively) are the short-cadence resolution of both missions. The dashed black line represent the induced TTVs for a ploonet with  $a=0.75$ au, in a circular orbit. It is included as reference.}
\label{fig:ttv}
}
\end{figure*}

\subsection{Noisy lightcurves}

To simulate the ploonet lightcurve, we decompose the transit of the object in three components: the ploonet (icy/volatile) core, the atmosphere and the cometary-like trail (labeled as \textit{coma} in Fig. \ref{fig:transit}) developed by the effect of the orbital motion and the action of the stellar wind. 

We assume that the projected images of the core and the atmosphere are circles with similar radius to those presented in Fig. \ref{fig:radius-evol}, for the case of a ploonet with albedo A = 0.89, at t = 1 Gyr.
The gas-trail is modeled with the projected area of a cone with a vertex in the ploonet position and a direction similar to the velocity vector.  We simulate the light curve of this cone-shaped tail as that produced by $500\,000$ meter-sized particles randomly distributed in the volume of the cone.  The inner radius of the cone is equal to radius of Titan (a moon of Saturn) while the outer radius is several times this quantity.

Particle sizes inside the coma are distributed according to a power law of slope $-1.5$. According to our evaporation model, a ploonet should lose $\sim 4 \times 10^9$ kg during the time of a single transit, resulting in $\sim 7000$ kg per simulated particle. This value was used to estimate the average particle size.

To obtain the lightcurve, we use the code \href{https://www.cfa.harvard.edu/~lkreidberg/batman/}{\tt Batman} \citep{Kreidberg2015}, assuming a simple linear model for the stellar limb-darkening, with coefficient $c=0.3$. 

The result of simulating the light curve of a typical ploonet are presented in the lower panel of Fig. \ref{fig:transit}.  The depth of th e transit is of the order of tens of ppm that could be detectable with present and future photometric surveys.  The characteristic shape is rather distinctive.  Since the probability that objects as small as a ploonet, actually have rings or moons, is very low, this shape could be a clear indication of a rapidly evaporating envelope.

\rev{Furthermore, using as input the calculated distribution of inclinations presented in \autoref{subsec:orbital}, the probability of a double transit (i.e. the transit of the planet followed by that of the noisy ploonet) was calculated. For each survivor ploonet, the planet impact parameter of its progenitor planet, $b_\mathrm{p}$, was randomly distributed in a uniform fashion between -1 and 1. Then, the inclination of the ploonets was transformed according to this new reference frame, and the impact parameter of the ploonets, $b_\mathrm{pl}$, was computed to look for simultaneous transits (i.e. $b_\mathrm{pl} \in [-1,1]$). We found that the likelihood for such an event is very small ($\sim\, 2$ per cent), which is possibly due to the assumed initial inclination of the system in the numerical integration.  
A system where the initial inclinations of moons are equal to zero (i.e. coplanar), should have a chance of double transit significantly higher, but a lower lifespan}.

\subsection{Transit Timing Variations}

\mhl{The close vicinity between the parent planet and a ploonet, can also trigger a second order effect in planetary lightcurves, namely Transit Timing Variations (TTV).  TTV are small deviations in the periodicity of the transit signal due to a close planetary companions.  They have been used to search for moons and other unseen planets.} 

\mhl{In the context of this work, both objects (the ploonet and its progenitor planet) should experience small deviations in the periodicity of their transits, being more notorious that of the ploonet (they are far lighter than the planet and their orbits could still be unstable). In the case of ploonets having tilted orbits the probability of observing mutual transit decrease, but its disturbing effects on the planet could be still notorious.}

\mhl{It is worth noting that TTVs produced by ploonets could appear as strong, ephemeral or even, single periodicity variation when very close-encounters happen between planet and ploonet. Moreover, in spite of the low mass of the ploonets, the fact of having similar semi-major axes improve their disturbing effect causing notorious deviations on its transit.}         

\mhl{Several authors (see e.g. \citealt{Nesvorny09,Vokrouhlick2014,Lillo-Box18} and references therein) have already studied the problem of the stability and detection by TTVs of sub planetary coorbital companions in extrasolar systems.} In this work, we adopted the prescription suggested in the guide of the N-body simulator \href{http://rebound.readthedocs.io/en/latest/index.html}{\tt REBOUND} \citep{Rein2012}.  We use the software to calculate the induced TTVs, for at least one member of each ploonet categories, ie. inner, outer, crossing, or coorbital (\autoref{subsec:orbital}). 
In Fig. \ref{fig:ttv}, we show the resulting TTV signal for the case simulated in \autoref{subsec:orbital}.

It is worth noting that cadences short-cadence capabilities of the \textit{Kepler'}s \citep{Gilliland2010} and TESS  photometry\footnote{\href{https://tess.mit.edu/science/observations/}{https://tess.mit.edu/science/observations/}} are below than the period deviations calculated for our systems, being the nearby object, as consequence, in principle, detectable.

\section{Summary and discussion}
\label{sec:summary}

From a dynamical and radiative perspective, the regions close to the star are very hostile for exomoons. In fact, the survival likelihood of a moon around a close-in planet is rather low (see \autoref{subsec:ejection}), as the probability of its detection. 
In this work, we have explored the dynamical and morphological evolution of a given population of ploonets, i.e. expelled exomoons. In addition, we computed the resulting observational signatures and found that ploonets are in principle detectable with the current observing capacities. 

In the scenario considered here, the exchange of angular momentum between the three bodies (star, planet, moon) leads to moon's unbound states with respect to the planet. Among all the ejected moons, the ones which survive over long time-scales (at least for several Myr) are of particular interest since they could potentially become planets, by accreting further material from the disc. In our realistic adiabatic model, moons are expelled without the requirement of any extra disturbing object, hence the phenomena discussed here could partially contribute to explain the lack of positive detection of exomoons (see also \citealt{Ramirez2018dm}). It is worth mentioning that there is an alternative channel to produce ploonets through particle-particle scattering, as shown by \citealt{Hong2018}.

Our results show that a subset of ploonets is able to survive for several \rev{hundreds of} Myrs orbiting close to their parent planet. However, a large fraction of these bodies ended either colliding with the planet or the star, or being ejected from the system. In the former case, low-energetic collisions with the planet occur because the planet and the ploonet share the same sense of orbital motion and have similar orbital energies. This is extremely relevant because these \textit{soft collisions} can lead to close-in and tilted exoring formation \citep{Sucerquia2017}. Interestingly, moon-star collisions could explain the anomalous spectroscopic features of the stars Kronos \& Krios  (HD 240430 and HD 240429) \citep{Oh2018}, which show deep traces of heavy elements. In fact, this suggests recent events of planetary cannibalism. 

Due to its vicinity with the star, a ploonet is subject to extreme processes such as surface sublimation and atmosphere evaporation. Both processes dramatically affect its morphology since the atmospheric envelope increases and evaporates, which translates into mass loss. Consequently, in the occurrence of transits, ploonets leave a characteristic signature in the lightcurve, which can be potentially observed. In fact, evaporating atmospheres from small bodies (or exocomets) have been already suggested from spectroscopic \citep{Iglesias2018} and photometric evidence  (see Section \ref{sec:intro}). These puzzling lightcurve profiles and spectra are readily similar to the ones obtained for evaporating ploonets (Fig. \ref{fig:transit}). 


\rev{An important outcome from our simplistic atmospheric model is that \textit{at the end} of a ploonet life, the size of its gaseous envelope increases dramatically, which makes the planet's effective cross-section to appear up to an order of magnitude larger when compared to its initial size. The combination of deeper transit light curves, and smaller induced radial-velocity measurements of such low-mass voluminous bodies, could be interpreted as two configurations that may occur even when the moon is still orbiting the planet: 1) a \textit{super-puff planet} in the mini-Neptunes regime, and 2) a massive or voluminous exomoon around a close-in giant planet \citep{Teachey2018}. The second system has faint associated TTV/TDV signals, since moon masses are small, and they are located on large semi-major axes as a result of their tidal evolution.}

In addition, since the planet and the ploonet orbits roughly coincide, ploonets can also affect the planetary light curve periodicity. Interestingly, as shown in Fig. \ref{fig:ttv}, \textit{Kepler} and TESS are sensitive enough to observe these TTV effects. Hence, combining both signatures, we should see non-periodical, deep and noisy planetary lightcurves caused by the ploonet and its expelled material. These  light curve features closely resemble the famous example of KIC-8462852, which exhibits a very noisy, non-periodical, and variable signal.

The detectability of a ploonet not only depends on its size, but also on its orbital distance from the planet. The slight variations on light curves (i.e. timings and duration) due to a close-in ploonet have observational limitations or are restricted to the proximity of the transit, shortening the time interval to detect its signal. This in turn reduces the scatter of TTV/TDV residuals and, affects the detection statistics regardless of the employed method. Throughout this endeavor there is another significant constraint to be ascertained: the transit configuration of the planet can also vary because of perturbations, e.g. starspots that can be dimmed by chance due to a single planet. To exclude any kind of perturbation as the origin of light-curve variations, the detected systemic changes on the shape of the transit and its vicinity have to independently be simulated and analysed. Whilst these calculations are feasible, it is not in the purview of this work.

To conclude, on the one hand, if exomoons collide with their host star/planet it is possible to produce some of the recent puzzling lightcurves observed. On the other hand, if they survive in co-rotation with the planet, then ploonets could eventually become small sized planets, be captured or collide with other planets of the system. This will be the subject of forthcoming research.

\section*{Acknowledgements} 

M.S. is supported by Colciencias (647) and the CODI/UdeA. J.A.M. acknowledges funding support from Macquarie University through the Macquarie University Research Excellence Scholarship (`iMQRES MRES').  J.I.Z. is supported by Vicerrector\'ia de Docencia UdeA. N.C. acknowledges financial support provided by FONDECYT grant 3170680. The authors thank Jorge Cuadra for helpful remarks and partial funding from CONICYT-Chile through FONDECYT (1141175) and Basal (PFB0609) grants. $N$-body computations were performed at Mulatona Cluster from CCAD - UNC, which is part of SNCAD - MinCyT, Argentina. The authors thank to the referee David Kipping for its valuable comments and many constructive suggestions. M.S Thanks also Professors G. Chaparro, A. Garc\'ia and J. Cuadra for their carefully reading of this work and the improvements suggested.    


%
\bibliographystyle{mnras}
\bibliography{references} 

\bsp	
\label{lastpage}
\end{document}